\documentclass[10pt]{article}
\usepackage[T1]{fontenc}
\usepackage{textcomp}
\usepackage{mathpazo}
\usepackage{authordate1-4}
\usepackage{graphicx}
\usepackage{amsmath,amssymb}
\begin{document}

\title{Live and Dead Nodes}
\date{}
\author{S.~Lehmann\thanks{Technical University of Denmark. Informatics and
        Mathematical Modeling, Building 321. DK-2800 Kgs. Lyngby, Denmark.}~~and
        A.~D.~Jackson\thanks{The Niels Bohr Institute, Blegdamsvej 17, DK-2100
        Copenhagen \O, Denmark}}
\maketitle

\section*{\emph{Abstract}}
\begin{small}
In this paper, we explore the consequences of a distinction between
`live' and `dead' network nodes; `live' nodes are able to acquire
new links whereas `dead' nodes are static. We develop an
analytically soluble growing network  model incorporating this
distinction and show that it can provide a quantitative description
of the empirical network composed of citations and references (in-
and out-links) between papers (nodes) in the SPIRES database of
scientific papers in high energy physics. We also demonstrate that
the death mechanism alone can result in power law degree
distributions for the resulting network.
\end{small}

\section{Introduction}
The study and modeling of complex networks has expanded rapidly in
the new millennium and is now firmly established as a science in its
own right \cite{watts:99,albert:02,dorogovtsev:02,newman:03a}. One
of the oldest examples of a large complex network is the network of
citations and references (in- and out-links) between scientific
papers (nodes)
\cite{price:65,redner:98,lehmann:03,lehmann:05,redner:04}. A very
successful model describing networks with power-law degree
distributions is based on the notion of \emph{preferential
attachment}. The principles underlying this model were first
introduced by Simon \cite{simon:57}, applied to citation networks by
de Solla Price \cite{price:76}\footnote{More precisely, de Solla
Price was the first person to re-think Simon's model and use it as a
basis of description for \emph{any} kind of network,
cf.~\cite{newman:03a}.}, and independently rediscovered by
Barab\'a{}si and Albert~\cite{barabasi:99}. Various modifications of
the preferential attachment model have appeared more recently. In
the present context, the key papers on preferential attachment are
\cite{lehmann:03,lehmann:05,krapivsky:00a,krapivsky:01,klemm:02}.
Simplicity is both the primary strength and the primary weakness of
the preferential attachment model. For example, preferential
attachment models tend to assume that networks are homogeneous. When
networks have significant and identifiable inhomogeneities (as is
the case for the citation network), the data can require
augmentation of the preferential attachment model to account for
them.

The primary conclusion of Ref.~\cite{lehmann:03} is that the
majority of nodes in a citation network `die' after a short time,
never to be cited again.  A small population of papers remains
`alive' and continues to be cited many years after publication. In
Ref.~\cite{lehmann:05} it was established that this distinction
between live and dead papers is an important inhomogeneity in the
citation network that is not accounted for by the simple
preferential attachment model. Interestingly, a similar distinction
between live and dead nodes was recently independently suggested by
\cite{redner:04}. In this paper, we will explore how the distinction
between live and dead papers manifests itself in network models and
thus suggest an extension of the preferential attachment model.

\section{The SPIRES data}\label{sec:data}
The work in this paper is based on data obtained from the
SPIRES\footnote{SPIRES is an acronym for `Stanford Physics
Information REtrieval System' and is the oldest computerized
database in the world. The SPIRES staff has been cataloguing all
significant papers in high energy physics and their lists of
references since 1974. The database is open to the public and can
be found at http://www.slac.stanford.edu/spires/.} database of
papers in high energy physics. More specifically, our dataset is
the network of all citable papers from the theory subfield, ultimo
October 2003. After filtering out all papers for which no
information of time of publication is available and removing all
references to papers not in SPIRES, a final network of $275\,665$
nodes and $3\,434\,175$ edges remains.

\begin{figure}[htbf]
\centering
  \includegraphics[width=\hsize]{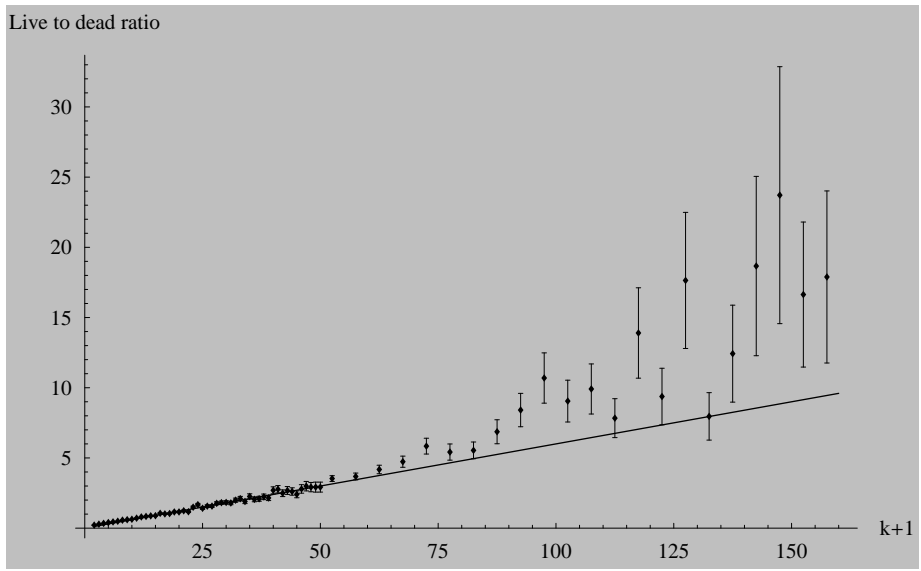}
  \caption{Displayed above is ratio of live to dead papers as a function of $k$.
  Error bars are calculated from square roots of the citation counts in each
  bin. Also, a straight line is present to illustrate the linear relationship between the
  live and dead populations for low values of $k$.
  }\label{fig:invratio}
\end{figure}

Above we described a dead node as one that no longer receives
citations, but how does one define a dead node in \emph{real} data?
We have tested several definitions, and the results are
qualitatively independent of the definition chosen. Therefore, we
can simply define live papers as papers cited in 2003.  While we
acknowledge the existence of papers that receive citations after a
long dormant period, such cases are rare and do not affect the large
scale statistics. In Figure~\ref{fig:livedead}, the (normalized)
degree distributions of live and dead papers in the SPIRES data are
plotted, and it is clear that the two distributions differ
significantly. Having isolated the dead papers, we are not only able
to plot them; we can also determine the empirical ratio of live to
dead  papers as a function of the number of citations per paper,
$k$. In Figure~\ref{fig:invratio} this ratio is displayed with $k$
ranging from $1$ to $150$ (Papers with zero citations are dead by
definition.) Over most of this range, the data is well described by
a straight line. Note that the data for dead papers with high
citation counts is very sparse. For example, only $0.15\%$ of the
dead papers have more than 100 citations, so the statistics beyond
this point are highly unreliable. More generally, a linear plot of
the ratio of live to dead papers provides a pessimistic
representation of the data.  We therefore conclude that the ratio of
\emph{dead to live} papers is relatively well described by the
simple form $1/(k+1)$ for all but the largest values of $k$, for
which the number of dead papers is overestimated by a factor of two
to three. In the following section, we will make use of this
relation to extend the preferential attachment model to include dead
nodes.

\section{The Model}\label{sec:model}
The basic elements of the preferential attachment model are
\emph{growth} and \emph{preferential attachment} \cite{barabasi:99}.
The simplest model starts out with a number of initial nodes and at
each update, a new node is added to the database. Each new node has
$m$ out-links that connect to the nodes already in the database.
Each new node enters with $k=0$ real in-links. This is the
\emph{growth} element of the model. Note that, since we have chosen
to eliminate all references to papers not in SPIRES from the
dataset, there is a sum rule such that the average number of
citations per paper is also $m$. \emph{Preferential attachment}
enters the model through the assumption that the probability for a
given node already in the database to receive one of the $m$ new
in-links is proportional to its current number of in-links. In order
for the newest nodes (with $k=0$ in-links) to be able to begin
attracting new citations, we load each node into the database with
$k_0=1$ `ghost' in-links that can be subtracted after running the
model.  The probability of acquiring new citations is proportional
to the \emph{total} number of in-links, both real and ghost
in-links.

One of the simplest ways to implement this simple incarnation of the
preferential attachment model described above is to regard $k_0$ as
a free parameter. This allows us to estimate when the effects of
preferential attachment become important. Since there is no \emph{a
priori} reason why a paper with 2 citations (in-links) should have a
significant advantage over a paper with 1 citation, it is preferable
to let the data decide. Thus, in our model, the probability that a
live paper with $k$ citations acquires a new citation at each time
step is proportional to $k+k_0$ with $k_0>0$. Also, note that we can
think of the displacement $k_0$ as a way to interpolate between full
preferential attachment ($k_0=1$) and no preferential attachment
($k_0 \to \infty$).

The significant  extension of the simple model to be considered here
is that, in our model, \emph{each paper has some probability of
dying at every time step}. From Section~\ref{sec:data}, we have a
very good idea of what this probability should be:
Figure~\ref{fig:invratio} shows us that for a paper with $k$
citations, this probability is proportional to $1/(k+1)$ to a
reasonable approximation. With this qualitative description of the
model in hand, we proceed to its solution.

\section{Rate Equations}
One very powerful method for solving preferential attachment network
models is the rate equation approach, introduced in the context of
networks by~\cite{krapivsky:00a}. Let $L_k$ and $D_k$ be the
respective probabilities of finding a live or a dead paper with $k$
real citations. As explained above, we load each paper into the
database with $k=0$ real citations and $m$ references. The rate
equations become
\begin{eqnarray}
    L_k &=& m(\lambda_{k-1}L_{k-1}-\lambda_{k}L_{k})-\eta_k L_k+\delta_{k,0}\label{eq:rateeqns}\\
    D_k &=& \eta_k L_k,\label{eq:rateeqns2}
\end{eqnarray}
where $\lambda_k$ and $\eta_k$ are rate constants. Since every paper
has a finite number of citations, the probabilities $L_k$ and $D_k$
become exactly zero for sufficiently large $k$; we also define $L_k$
to be zero for $k<0$. In this way, all sums can run from $k=0$ to
infinity. These equations trivially satisfy the normalization
condition
\begin{equation}\label{eq:totnorm}
    \sum_k (L_k+D_k)=1,
\end{equation}
for any choice of $\eta_k$ and $\lambda_k$. However, we also demand
that the mean number of references  is equal to the mean number of
papers
\begin{equation}\label{eq:meannorm}
    \sum_k k(L_k+D_k)=m.
\end{equation}
This constraint must be imposed by an overall scaling of $\eta_k$
and $\lambda_k$. The model described in Section~\ref{sec:model}
corresponds to a choice of $\eta_k$ and $\lambda_k$, where
\begin{equation}\label{eq:lambdadef}
    m\lambda_k = a(k+k_0)
\end{equation}
is the preferential attachment term and
\begin{equation}\label{eq:etadef}
    \eta_k=\frac{b}{k+1}
\end{equation}
corresponds to the previously described death mechanism. We insert
Equations~(\ref{eq:lambdadef}) and~(\ref{eq:etadef}) into
Equation~(\ref{eq:rateeqns}) and perform the recursion to find
\begin{equation}
    L_k = \frac{\Gamma(k+2)}{a k_1 k_2}\frac{\Gamma(k+k_0)}{\Gamma(k_0)}
    \frac{\Gamma(1-k_1)}{\Gamma(k-k_1+1)}\frac{\Gamma(1-k_2)}{\Gamma(k-k_2+1)}\label{eq:finalsolution},
\end{equation}
and of course $D_k = b L_k/(k+1)$. The two new constants, $k_1$ and
$k_2$ are solutions to the quadratic equation
\begin{equation}\label{eq:quadratic}
    (a(k+k_0)+1)(k+1)+b=0
\end{equation}
as a function of $k$.

\section{The $k_0 \to \infty$ Limit}
Before moving on, let us explore the limit where $k_0 \to \infty$
and preferential attachment is turned off. In this regime, the
network is, of course, completely dominated by the death mechanism.
We can either obtain this limit by again solving
Equations~(\ref{eq:rateeqns}) and~(\ref{eq:rateeqns2}) with
$\lambda_k = constant$ and $\eta_k=b/(k+1)$, or we can make the more
elegant replacement $\alpha = ak_0$ in
Equation~(\ref{eq:finalsolution}), and then take the limit $k_0 \to
\infty$ for fixed $\alpha$. The two approaches are equivalent. We
find
\begin{equation}\label{eq:k0inftysolution}
    L_k = \frac{1}{\alpha}
    \left( \frac{\alpha}{1+\alpha} \right)^{k+1}
    \frac{(\frac{b}{1+\alpha})!(k+1)!}{(\frac{b}{1+\alpha}+k+1)!},
\end{equation}
and the $D_k$ are still simply $b L_k/(k+1)$. With this expression
for $L_k$, let us consider the limit of $\alpha \to \infty$ and $b
\to \infty$ with the ratio $r=b/(\alpha+1)\approx b/\alpha$ fixed.
In this limit, it is tempting to replace the term $\alpha/(\alpha +
1)$ by one\footnote{For present purposes, this is appropriate when
$r \ge 2$. When $r < 2$, the neglected factor is essential for
ensuring the convergence of the average number of citations for the
live and dead papers $m_L$ and $m_D$.}. In this case, the use of
identities, such as
\begin{equation}\label{eq:sumidentity}
    \sum^\infty_{k=1} \frac{k!}{(k+r)!} =
    \frac{1}{(1-r)r!}
\end{equation}
enable us to compute the fraction of dead papers $f$, and the
average numbers of citations for live and dead papers. The results
are simply
\begin{eqnarray}
1 - f  &=& \frac{1}{\alpha-1}\\
m_L &=& \frac{2}{r-2}\\
m_D &=& \frac{1}{r-1},
\end{eqnarray}
and the average number of citations for all papers is evidently $m
= (1-f)m_L + f m_D$. The fraction of dead papers is $f\to 1-
\mathcal{O}(1/b)$ and the average number of citations for all
papers approaches $m_D$.

The most important result, however, is that in this limit we find
that
\begin{equation}\label{eq:powerlaws}
    L_k \sim \frac{1}{k^r}
    \quad \textrm{and} \quad
    D_k \sim \frac{b}{k^{r+1}},
\end{equation}
where we assume that $k>r$. Thus, \emph{we see that power law
distributions for both live and dead papers emerge naturally in the
limit of $f \to 1$}. In the literature, power laws in the degree
distributions of networks are often regarded as an indication that
preferential attachment has played an essential part in the
generation of the network in question.  It is thus of considerable
interest to see an alternative and quite different way of obtaining
them.

\section{The Full Model}
\begin{figure}[htbf]
\centering
  \includegraphics[width=\hsize]{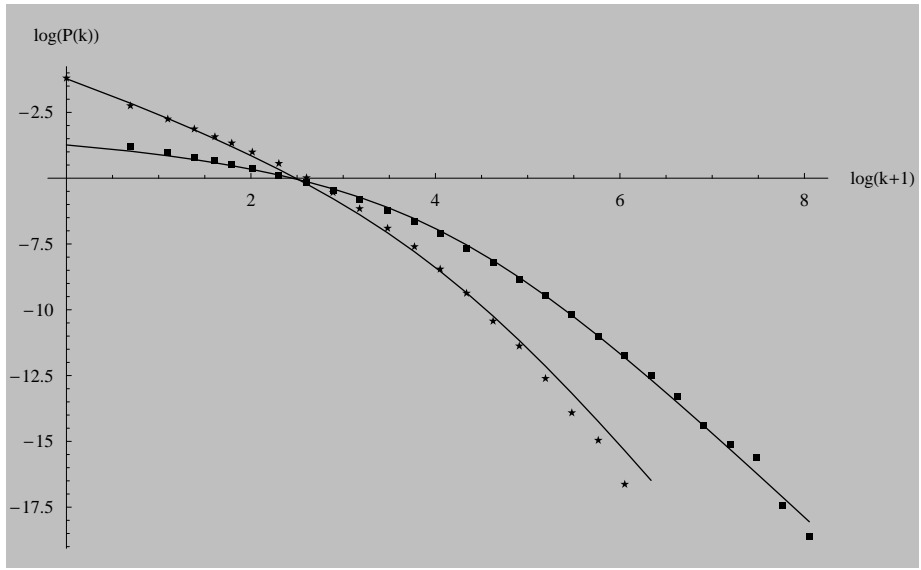}
  \caption{Log-log plots of the normalized degree distributions of live and
  dead papers. The filled squares represent the live data and the
  stars represent the dead data. Both lines are the result of a fit to the
  live data (filled squares) alone.
  }\label{fig:livedead}
\end{figure}

\begin{figure}[htbf]
\centering
  \includegraphics[width=\hsize]{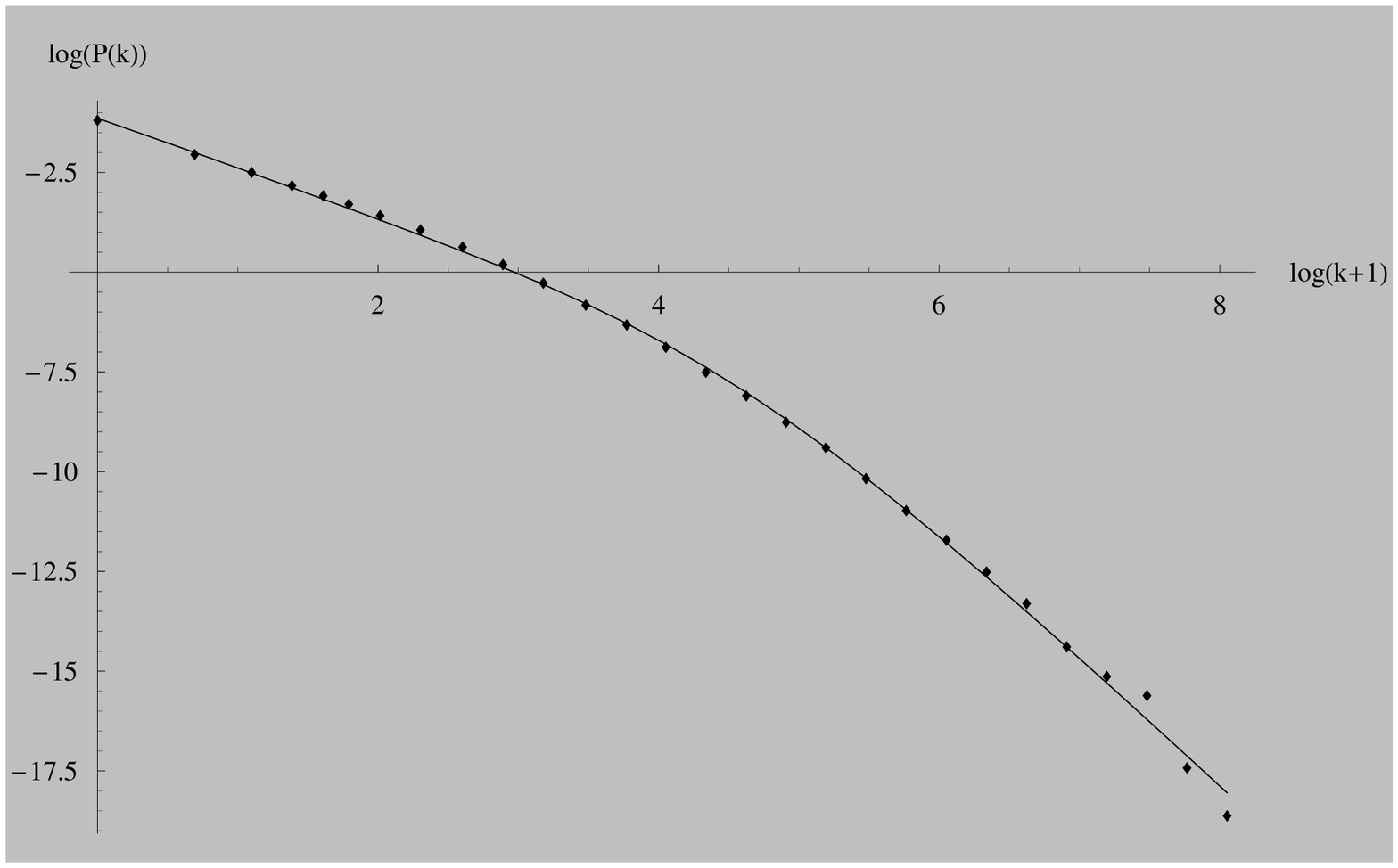}
  \caption{A log-log plot of the normalized degree distribution of
  all papers (live plus dead). The points are the data;
  the fit (solid line) is derived from the fit to the live papers (filled squares) in
  Figure~\ref{fig:livedead}.
  }\label{fig:total}
\end{figure}

Let us now return to the full model and see how it compares to the
data from SPIRES. With all zero cited papers in the dead category,
the data yields the following average values: $m_L=34.1$, $m_D=4.5$
and $m=12.8$. The fraction of live papers is $f=27.0\%$. With an
rms.~error of only 21\%, we can do a least squares fit of $L_k$ to
the distribution of live papers with parameters $k_0 = 65.6$,
$a=0.436$, and $b = 12.4$. Although only the live data (the squares
in Figure~\ref{fig:livedead}) is fitted, the agreement with the
empirical data in Figures~\ref{fig:livedead} and~\ref{fig:total} is
quite striking.

From the model parameters $k_0, a, b$, we can calculate mean
citation numbers for the fit of $32.9$, $4.25$, and $12.8$ for the
live, dead, and total population respectively; the fraction of live
papers is found to be $29.8 \%$. More interestingly, we learn from
the fit that $7.5\%$ of the papers with 0 citations \emph{are
actually alive}. If we assign this fraction of the zero-cited papers
to the live population, we find the following corrected values for
the average values $31.5$, $4.6$ and $12.5$ for the live, dead, and
total population respectively; the fraction of live papers is
adjusted to  become $29.2 \%$. Again, this is a striking agreement
with the data. There is so little strain in the fit that we could
have determined the model parameters from the empirical values of
$m_L$, $m_D$, and $f$. Doing this yields only small changes in the
model parameters and results in a description of comparable quality!

Figure~\ref{fig:livedead} reveals that fitting to the live
distributions, results in systematic errors for high values of $k$
when we extend the fit to describe the dead papers, but this is not
surprising. Recall the similarly systematic deviations from the
straight line seen in Figure~\ref{fig:invratio}. This figure also
explains why the fit to the total distribution shows no deviations
from the fit for high $k$-values even though the total fit includes
both live and dead papers---live papers dominate the total
distribution in this regime. The obvious way to fix this problem is
via a small modification of the $\eta_k$. In summary, the full model
is able to fit the distributions of both live and dead papers with
remarkable accuracy.

One drawback, with regard to the full solution is the relatively
impenetrable expression for $L_k$ in
Equation~(\ref{eq:finalsolution})---associating any kind of
intuition to the conglomerate of gamma-functions presented there can
be difficult. Let us therefore demonstrate that $L_k$ can be well
approximated by a two power law structure. We begin by noting that,
in the limit of large $k_0$ (as it is the case here), the values of
$k_1$ and $k_2$ are simply
\begin{eqnarray}
k_1 &=& - \frac{1}{a} + \frac{b}{a k_0} - k0 \label{eq:k1largek} \\
k_2 &=& -1-\frac{b}{a k0}. \label{eq:k2largek}
\end{eqnarray}
Now, let us write out only the $k$-dependent terms in
Equation~(\ref{eq:finalsolution}) and assign the remaining terms to
a constant, $C$
\begin{eqnarray}
L_k &=& C \frac{(k+k_0-1)!}{(k-k_1)!} \frac{(k+1)!}{(k-k_2)!}\\
    &\approx& C\frac{1}{(k+k_0-1)^{1-k_0-k_1}}\frac{1}{(k+1)^{-(1+k_2)}}\label{eq:approximation}\\
    &\approx& C \frac{1}{(k+k_0-1)^{1+\frac{1}{a}-\frac{b}{a k_0}}}\frac{1}{(k+1)^{\frac{b}{a k_0}}}\label{eq:approximation2},
\end{eqnarray}
In Equation~(\ref{eq:approximation}), we have utilized the fact
that
\begin{equation}\label{eq:relation}
    \frac{(x+s)!}{x!} \approx x^s
\end{equation}
when $x\to\infty$, and in Equation~(\ref{eq:approximation2}) we
have inserted the asymptotic forms of $k_1$ and $k_2$, from
Equations~(\ref{eq:k1largek}) and~(\ref{eq:k2largek}).

This expression for $L_k$ in Equation~(\ref{eq:approximation2}) is
only valid for large $k$ and $k_0$, but it proves to be remarkably
accurate even for smaller values of $k$. With the asymptotic forms
of $k_1$ and $k_2$ inserted, we can explicitly see that the first
power law is largely due to preferential attachment and that the
second power law is exclusively due to the death mechanism. The form
for very large $k$ is unaltered by the parameter $b$. This is not
surprising, since there is a low probability for highly cited papers
to die. We see that the primary role of the death mechanism in the
full model is to add a little extra structure to the $L_k$ for small
$k$.

\section{Conclusions}
Compelled by a significant inhomogeneity in the data, we have
created a model that provides an excellent description of the SPIRES
database. It is obvious that the death mechanism $(b\ne 0)$ is
essential for describing the live and dead populations separately,
but less clear that it is indispensable when it comes to the total
data. Fitting the total distribution with a preferential attachment
only model $(b=0)$ results in $a=0.528$ and $k_0=13.22$ and with a
rms.~fractional error of $33.6\%$. This fit displays systematic
deviations from the data, but considering that the fit ignores
important correlations in the dataset, the overall quality is rather
high. The important lesson to learn from the work in this paper, is
that even a high quality fit to the global network distributions is
not necessarily an indication of the absence of additional
correlations in the data.

The most significant difference between the full live-dead model and
the model described above is expressed in the value of the parameter
$k_0$. The value of this parameter changes by a factor of
approximately 5, from $65.6$ to $13.2$. It strikes us as natural
that preferential attachment will not be important until a paper is
sufficiently visible for authors to cite it without reading it. We
thus believe that $k_0 \approx 66$ is a more intuitively appealing
value for the onset of preferential attachment. However, independent
of which value of the $k_0$ parameter one prefers, the comparison of
these two models clearly demonstrates the danger of assigning
physical meaning to even the most physically motivated parameters if
a network contains unidentified correlations or if known
correlations are neglected in the modeling process. Specifically, it
would be ill advised to draw strong conclusions about the onset of
preferential attachment if the death mechanism is not included in
the model making.

In summary, the live and dead papers in the SPIRES database
constitute distributions with significantly different statistical
properties. We have constructed a model which includes modified
preferential attachment and the death of nodes. This model is
quantitatively successful in describing the citation distributions
for live and dead papers.  The resulting model has also been shown
to produce a two power law structure. This structure provides an
appealing link to the work in \cite{lehmann:03}, where a two power
law structure was adopted to characterize the form of the SPIRES
data without any theoretical support.  Finally, we have been shown
that even in the absence of preferential attachment, the death
mechanism alone can result in power laws. Since many real world
networks have a large number of inactive nodes and only a small
fraction of active nodes, we are confident that this mechanism will
find more general use.

\section*{\emph{Acknowledgements}}
Our grateful thanks to T.~C.~Brooks at SPIRES without whose
thoughtful help we would have lacked all of the data!

\newpage

\newpage
\newpage

\end{document}